\UseRawInputEncoding
\documentclass[referee]{raa}           
\pdfoutput=1
\usepackage{graphicx,times}            
\usepackage{natbib}
\usepackage{amssymb,amsmath}
\usepackage{graphicx}
\usepackage{multirow}
\usepackage{longtable}
\usepackage{graphicx}
\usepackage{lscape}
\bibpunct{(}{)}{;}{a}{}{,}

\usepackage[pagebackref=true]{hyperref}

\begin{document}

  \title{Evolution of Galaxy Types and HI Gas in Hickson Compact Groups
}

   \volnopage{Vol.0 (20xx) No.0, 000--000}      
   \setcounter{page}{1}          

   \author{Yao Liu 
      \inst{1,2,3}
   \and Ming Zhu
      \inst{1,2}
   }

   \institute{National Astronomical Observatories, Chinese Academy of Sciences,
             Beijing 100012, China; {\it liuyao@nao.cas.cn}\\
        \and
             CAS Key Laboratory of FAST, NAOC, Chinese Academy of Sciences\\
        \and
             University of Chinese Academy of Science, Beijing 1000101, China\\
\vs\no
   {\small Received 20xx month day; accepted 20xx month day}}

\abstract{ Compact groups have high galaxy densities and low velocity dispersions, and their group members have experienced numerous and
frequent interactions during their lifetimes. They provide a unique environment to study the evolution of galaxies. We examined the galaxies types and HI contents in groups to make a study on the galaxy evolution in compact groups. We used the group crossing time as an age indicator for galaxy groups. Our sample is derived from the Hickson Compact Group catalog. We obtained group morphology data from the Hyper-Leda database and the IR classification based on Wide-Field Infrared Survey Explorer (WISE) fluxes from \cite{zucker2016hierarchical}. By cross-matching the latest released ALFALFA 100$\%$ HI source catalog and supplemented by data found in literature, we obtained 40 galaxy groups with HI data available. We confirmed that the weak correlation between HI mass fraction and group crossing time found by \cite{ai2018evolution} in SDSS groups also exists in compact groups. We also found that the group spiral galaxy fraction is correlated with the group crossing time, but the actively star-forming galaxy fraction is not correlated with the group crossing time. These results seem to fit with the hypothesis that the sequential acquisition of neighbors from surrounding larger-scale structures has affected the morphology transition and star formation efficiency in compact groups. 
}

   \authorrunning{Yao Liu \& Ming Zhu }            
   \titlerunning{Galaxy Types and HI Gas in HCGs }  

   \maketitle

%
%
\section{Introduction}           
\label{sect:intro}

Galaxies, made up of stars, dust, dark matter, and gas, are a fundamental component of the Large Scale Structure. Galaxies lie in the center of the dark matter halo and can clump together when dark matter halos grow over cosmic time through merging. Observations show that half to two-thirds of galaxies reside in group systems whose members ranges from a few to dozens (\citealt{huchra1982groups, berlind2006percolation, tempel2012groups}). It is believed that galaxy groups are where galaxies spent the majority of their life time. (\citealt{osmond2004gems, forbes2006group, freeland2009h, balogh2009colour}).

One of the key issue in galaxy evolution is the impact of galaxy environment, as measured by the density of galaxies per $\ Mpc^{3}$, on galaxy properties such as star formation rates, optical colors, and morphologies. For example, \cite{butcher1978apj} observed that the fraction of blue galaxies in galaxy clusters is higher at high redshift than at low redshift, implying more active star formation in the past, and the average stellar population is older in the present day. \cite{dressler1980catalog} found that late-type galaxies are preferentially found in low-density environments, while early-type galaxies populate high-density environments. However, these observations only give us a general picture on the galaxy evolution, the specific evolution process, such as how galaxies are drawn into groups and clusters is still unclear.

Neutral hydrogen (HI), observed with the HI 21 cm line, is an important component of galaxies which makes up roughly 50 $\%$  of the gas in the interstellar medium (ISM) of typical Milky Way-like spiral galaxies. Since HI gas is the fuel for star formation and is easily disturbed by environment effects, HI content in galaxies provides a tracer of intergalactic interaction. Many previous works focus on measuring the gas content of galaxies in clusters ( e.g. Virgo, \citealt{chung2009vla}; Coma, \citealt{gavazzi2006h}), or in groups (e.g. \citealt{pisano2007hi, kern2008h, kilborn2009southern}). These studies show that galaxy groups and clusters are deficient in HI (\citealt{haynes1984influence, verdes2001neutral, solanes2004hi, taylor2012arecibo}) and HI mass distribution in galaxies likely evolves with the environment (\citealt{springob2005morphology, kilborn2009southern}). Several physical processes have been suggested to explain how environments drive galaxy evolution such as ram pressure stripping caused by hydrodynamic interaction between ISM of galaxies and the intra-cluster medium (\citealt{gunn1972infall}), galaxy harassment caused by fast, minor gravitational encounters between galaxies, tidal interaction and mergers in low-velocity dispersion groups (e.g. \citealt{hibbard1995dynamical, freeland2009h, kern2008h}) and strangulation and evaporation in X-ray bright groups (\citealt{cowie1977thermal, rasmussen2008galaxy}). However, the detailed physical processes for HI deficiency in galaxy groups remain unclear.

Compact groups of galaxies (CGs) are groups with high number density (separated by a few galaxy radii), and low-velocity dispersion (radial median of roughly $\rm 200\;km\;s^{-1}$). These conditions favor interactions, and even mergers, making CGs an ideal laboratory for studying the physical processes in galaxy evolution. CGs are also found to be deficient in HI and the atomic gas in some groups shows complex structures and even spatial offsets from their host galaxies (\citealt{williams1987neutral, huchtmeier1997hi}).
In studies by \cite{verdes2001neutral,walker2012examining,cluver2013enhanced,alatalo2014catching,bitsakis2016studying,lisenfeld2017role}, the authors proposed an evolutionary scenario for CGs based on the analysis of the total HI content of Hickson compact groups(HCG). According to these authors, in an early stage, galaxies in CG lose their atomic gas in the outer parts of galaxies while molecular gas and dust are not much depleted. As the group evolves, the tidal interaction between galaxies and interactions with the intra-group gas will increase, which will affect the molecular gas content and star formation efficiency, and the tidally-stripped HI structures disperses into faint HI medium gradually. They suggested that the distribution and nature of HI in the tidal structure can be used to trace the evolution phase of the group. 

Another observational fact is that galaxies in CGs show a bi-modal distribution in mid-infrared (mid-IR) color space, and the dearth of canyon galaxies between active galaxies and quiescent galaxies has been interpreted as a rapid transition from actively star-forming to quiescent systems. (\citealt{walker2013optical, zucker2016hierarchical}).

To make a preliminary study on the galaxy evolution in CGs, we investigated the gas content of galaxies in a subsample of the HCG catalog. \cite{ai2018evolution} made a full census of the gas content of a group sample selected from the SDSS group catalog by cross-matching them with the ALFALFA 70 $\%$ HI source catalog and found a weak correlation between the group HI mass fraction and the group crossing time, suggesting that group crossing time is a good indicator of the group’s age. Nevertheless, by comparing mocked galaxy catalogs derived from the semi-analytic galaxy catalogs with observed CGs, researchers found that there exist clear differences between semi-analytic models and observed CGs and concluded that CGs should constitute a specific class of groups with a distinct evolutionary path (e.g. \citealt{snaith2011comparison, farhang2017evolution}). It is of interest to know whether the correlation found by \cite{ai2018evolution} still exists in CGs. In this paper, we study the relationship between the HI mass fraction and group crossing time. Also, we check the correlation between the group type fraction and other group properties, including group crossing time and HI content.
  Throughout the paper we use the Hubble constant $H_0$ = 70 $\ km\;s^{-1} Mpc^{-1}$.

\section{Sample and data}
\label{sect:data}

\subsection{Galaxies group catalogs}
We draw our sample from galaxy groups in the HCG catalog, which was first published by \cite{hickson1982systematic} based on systematic visual search of the Palomar Observatory Sky Survey prints, defined by the following criteria:
\begin{enumerate}
    \item N $\ge$ 4                           (population)
    \item $\theta _{N} \ge 3\theta _{G} $     (isolation)
    \item $\bar\mu _{G} < 26.0$               (compactness)
\end{enumerate}
Where N is the total number of galaxies within 3 mag of the brightest member, $\bar\mu _{G} $ is the total magnitude of these galaxies per $\ arcsec^{2}$ averaged over the smallest circle (angular diameter $\theta _{G} $ ) that contains their geometric centers, and $\theta _{N} $ is the angular diameter of the largest concentric circle that contains no other (external) galaxies within this magnitude range. Based on new radial velocity data from follow-up spectroscopic observations, \cite{hickson1992dynamical} excluded unrelated group members and reduced the HGC sample to 92 groups. To ensure the reliability of our investigation, we use the number of galaxy group members n $\ge$ 4 as data selection criteria. This is because n=4 will greatly reduce the standard deviation of group properties compared to n=3. For example, when calculating the median length of the two-dimensional galaxy-galaxy separation vector R, n=4 groups have 6 data points, twice as much as n=3. Further, We removed HCG 47,48 and 88 because they have estimated errors (Equation (1) in \citealt{hickson1992dynamical}) larger than the observed velocity dispersion, and so the intrinsic velocity dispersion cannot be determined. HCG 54, which consists of small optical knots ($\sim $0.6 $\ kpc$ diameter) embedded in a single 12 kpc-diameter cloud with a long tail about 20 $\ kpc$ long, was also excluded because it might be knots of at most two irregular galaxies (\citealt{verdes2001neutral}). Our investigation with HCGs also shows that the properties of HCG 54 have a serious deviation from the rest groups suggesting that it might be a false group. Finally, we got a sample containing 292 galaxies in 64 groups with n $\ge$ 4. There are 23 groups with n=3 are not included in our sample.

\subsection{Galaxy type and Morphology}
\cite{zucker2016hierarchical} derived WISE fluxes for a sample of 652 galaxies in 163 compact groups. Based on WISE mid-IR color plots ( $\log\left [  f_{12} /f_{4.6}  \right ] vs.\log\left [ f_{22}/f_{3.4}   \right ] $ ), they classify the galaxies into 3 classes: mid-IR active, mid-IR quiescent and mid-IR canyon which represents different levels of star formation. 93 galaxy groups of this catalog are from the Hickson Compact Group catalog (HCG). The other groups in Zucker’s catalogs are from the Redshift Survey Compact Group catalog (RSCG, \citealt{barton1996compact}) selected with the same criteria as that of the HCGs. We only used the HCG counterpart because we don’t have group crossing time for RSCGs. All our HCG groups and 88.94 $\%$ (243) of group members are classified with the mid-IR colors, and 35.27 $\%$(103) member galaxies in 48 (75 $\%$) groups are classified as actively star-forming galaxies.

Each galaxy in our sample has morphological data from the Hyper-Leda database (\citealt{paturel2003hyperleda}), which is available on the website http://leda.univ-lyon1.fr/leda/. They used an optimal De Vaucouleurs number for each galaxy to represent morphology. We use the same classification scheme as that in \cite{zucker2016hierarchical} shown in Table~\ref{Tab1}. In our research, a cut-off value of 0.5 was used to distinguish between spiral galaxies and elliptical galaxies. As a result, 98(33.56 $\%$) galaxies in 53(82.8$\%$) groups are classified as spirals.

%
\begin{table}
\begin{center}
\caption[]{ The Morphology Classification Scheme of De Vaucoleurs Number For Galaxies.}\label{Tab1}


 \begin{tabular}{clcl}
  \hline\noalign{\smallskip}
Vaucouleurs number & Morphogy    \\ \hline
-5$\sim$-3.5       & E           \\
-3.5$\sim$-2.5     & E$\sim$S0   \\
-2.5$\sim$-1.5     & S0          \\
-1.5$\sim$0.5      & S0$\sim$Sa  \\
0.5$\sim$2.5       & Sa$\sim$Sab \\
2.5$\sim$4.5       & Sb$\sim$Sbc \\
4.5$\sim$7.5       & Sc$\sim$Sd  \\
7.5$\sim$10.0      & Irr,Sdm     \\ \hline
\end{tabular}
\end{center}
\end{table}

\subsection{HI mass in HCGs}
\subsubsection{HI sources associated with HCGs}
We have investigated all existing HI data in the literature for our HCG subsample (292 galaxies in 64 groups), including the latest released ALFAFA 100 ($\alpha.100$) catalog (\citealt{haynes2018arecibo}), and the Springob/Cornell HI data catalog which, collected by \cite{springob2005morphology}, contains HI linewidth parameters and profiles for 8844 galaxies in the local universe observed by a variety of large single-dish radio telescopes (Arecibo, Effelsberg, Green Bank 91m, Green Bank 42m and Nancy). We also include the direct HI measurements of HCGs (\citealt{williams1987neutral,huchtmeier1997hi,borthakur2010detection}
) from the Arecibo radio telescope, the Green Bank radio telescope and the 100m Eeffelsberg antenna. 

The $\alpha.100$ catalog is available on the website http://egg.astro.cornell.edu/alfafa/data/, while the Springob/Cornell catalog can be searched from The Extragalactic Distance Database (EDD). Both of these two catalogs provide optical counterparts for the detected HI sources. We use the software TOPCAT to cross-match our HCG sample with the above two HI data catalogs using each galaxy’s Principal Galaxies Catalog (PGC) number in the Lyon Extragalactic Database (LEDA). After integrating the above data with single dish HI studies of HCGs from the other three, we selected HI data based on signal to noise ratio (SNR) when there are groups with duplicate measurements. We have double checked that these duplicated measurements are generally in good agreement with each other. In addition, We adopt HI data of HCG 16 from \cite{jones2019evolution}, who re-reduced HI raw data of HCG 16 mapped with the Very Large Array (VLA) in C and D configurations in 1991 and 1989 respectively, and studied them in detail. Finally, we obtained 40 galaxy groups accounting for 62.5$\%$ of the total group sample (64). In this subsample with HI data available, 159(86.41 $\%$) member galaxies have mid-IR colors, and 78 (42.40 $\%$) member galaxies in 36 (90 $\%$) groups are actively star-forming, while
81(44.02$\%$) member galaxies in 39 (97.50 $\%$) groups are classified as spirals. Comparing to the 64 HCG groups sample, this subsample appears to be both actively star-forming and spiral richer.

\subsubsection{HI source confusion and HI mass estimates}
\begin{figure}[ht]
  \begin{minipage}[t]{0.495\linewidth}
  \centering
   \includegraphics[width=70mm,height=60mm]{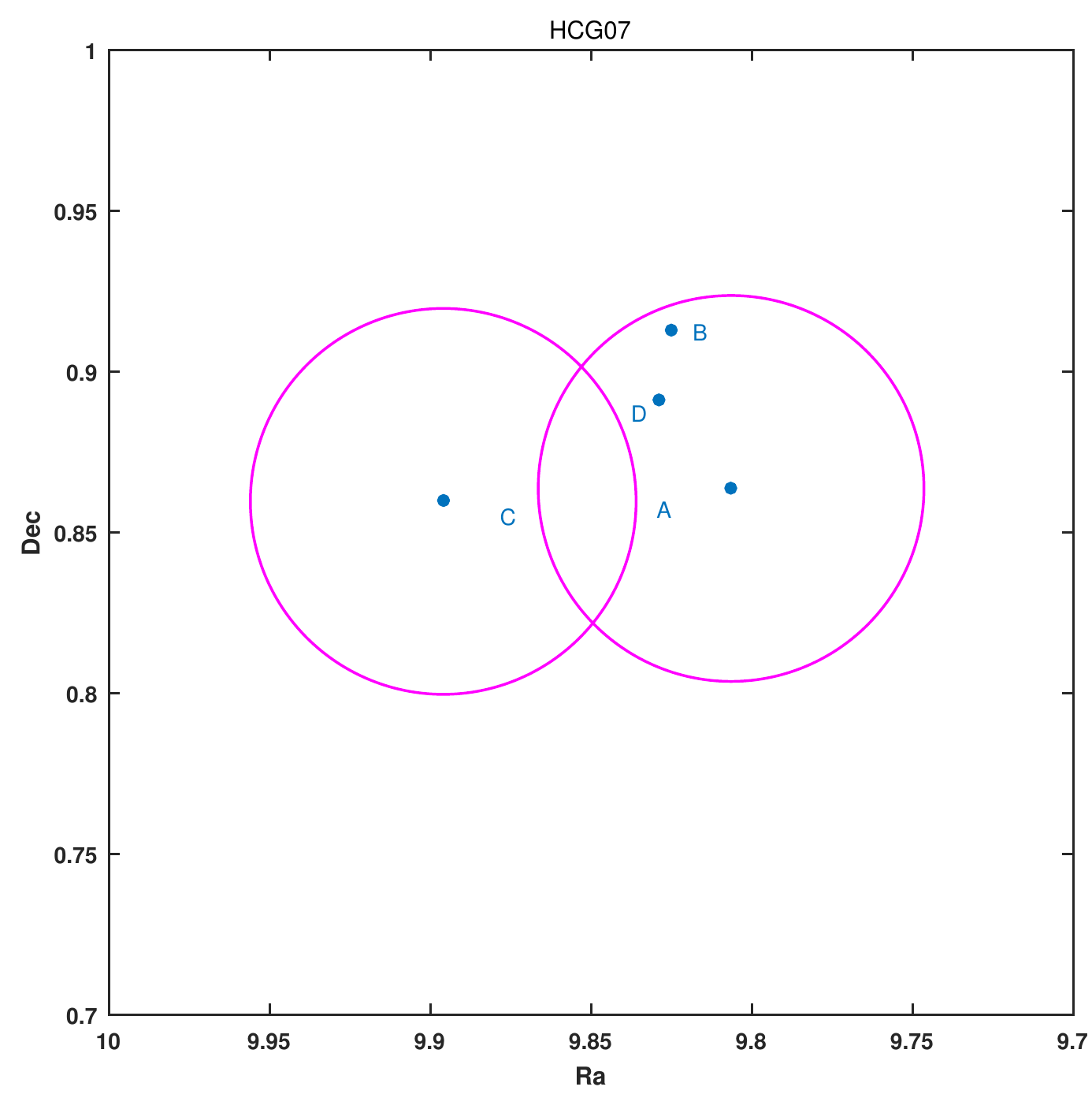}
  \end{minipage}%
  \begin{minipage}[t]{0.495\textwidth}
  \centering
   \includegraphics[width=70mm,height=60mm]{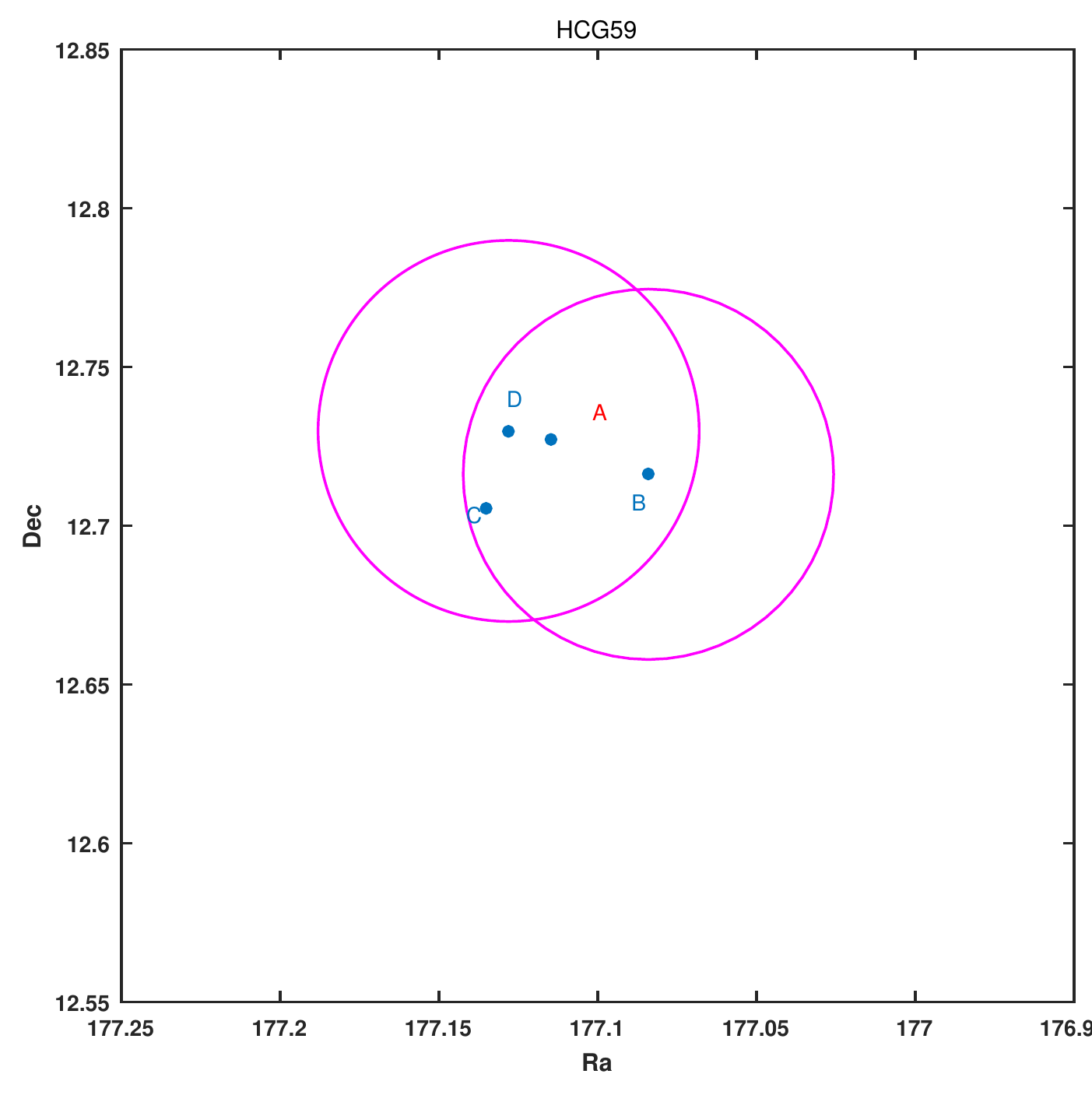}
  \end{minipage}%
  
    \begin{minipage}[t]{0.495\linewidth}
  \centering
   \includegraphics[width=70mm,height=60mm]{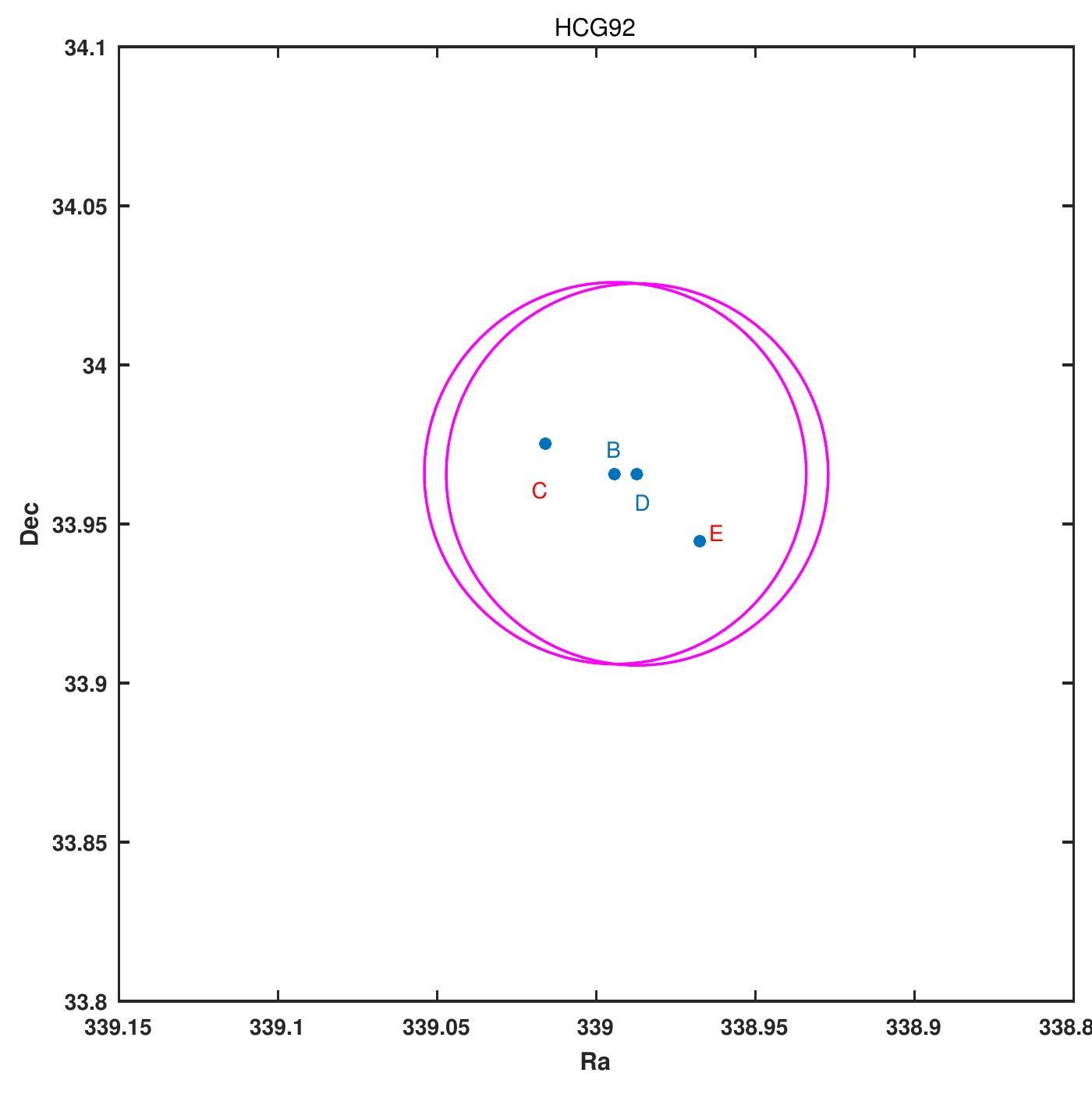}
  \end{minipage}%
  \begin{minipage}[t]{0.495\textwidth}
  \centering
   \includegraphics[width=70mm,height=60mm]{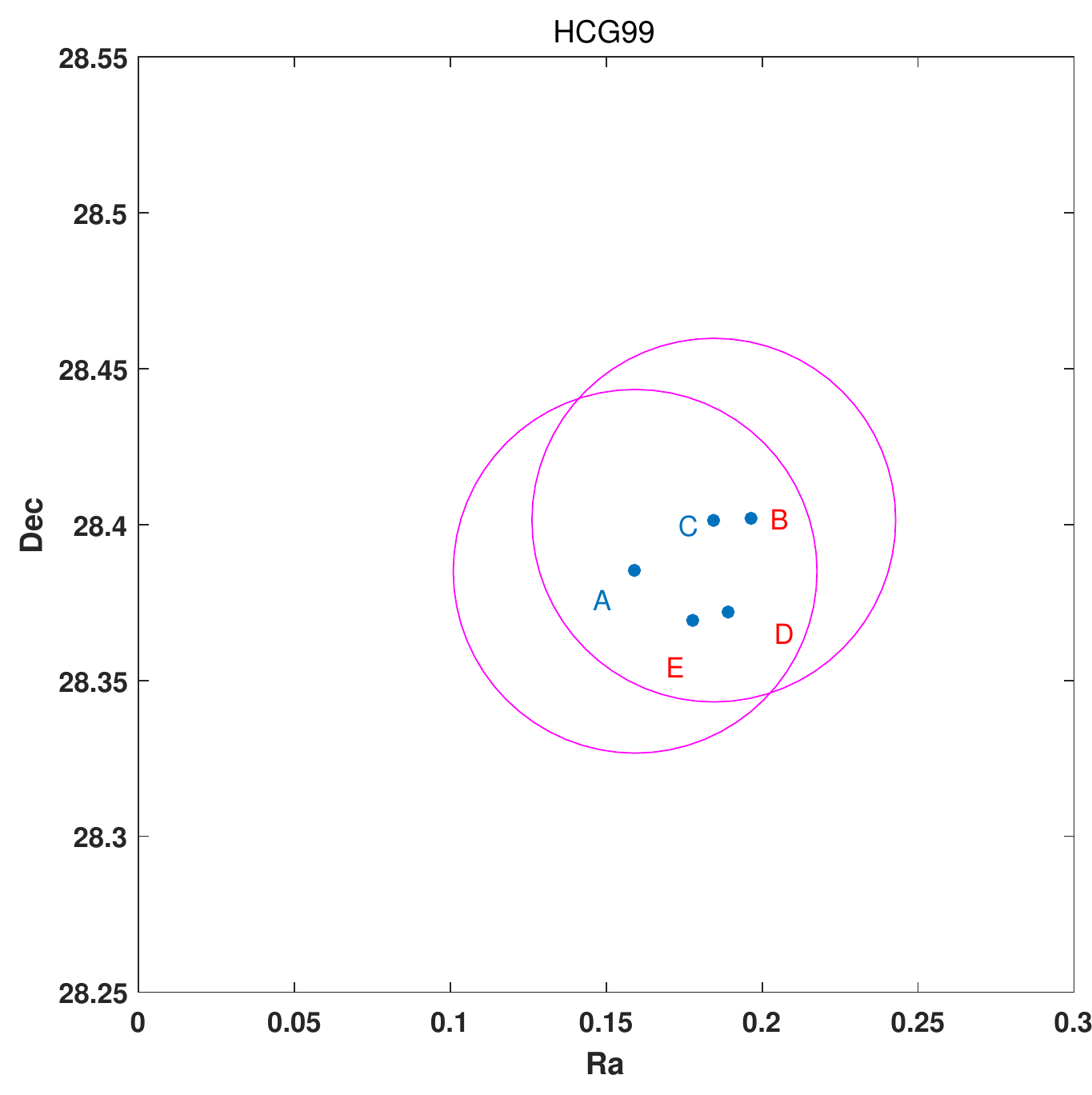}
  \end{minipage}%
  \caption{The ALFALFA beam coverage of HCG 7, 59, 92 and 99. Galaxies with observed HI data are marked with blue letter and galaxies having no detection are marked with red letter. We assumed the center of each circle is the HI source galaxy.}
  \label{Fig1}
\end{figure}

Since in compact groups, distance between galaxies could be too close compared to the Arecibo beam size in some cases, we need to cheek the confusion problem to ensure the accuracy of data. In our sample, there are 8 galaxy groups have more than one HI measurements which could be subject to confusion. They are HCG 7, 10 ,15 ,44, 54, 58, 92 and 99.
The beam size of ALFALFA is $3.5\times 3.8^\prime$  with an average $18^{\prime \prime}$ pointing error. ALFALFA beams can resolve the individual galaxies in HCG 10, 15, 44 and 58, thus would not have confusion problems. For these groups we can simply sum up each member galaxy’s HI flux to obtain the total group HI flux. For other 4 groups that could have confusion problem, e.g., HCG 7, 54, 92, and 99, we plot out the ALFALFA beams coverage in Figure~\ref{Fig1}. We only plot the Arecibo beams for the two most massive galaxies in each group. For HCG 7, the mean angular distance between HCG 7a, HCG 7d, and HCG 7b is $2^\prime 11^{\prime \prime}$. Thus the maximum value of the HI flux for these three galaxies should represent the total flux of them. We can use this flux and combine with the HI flux for HCG 7c to derive the total HI flux for HCG 7. The HI mass of HCG 7 (9.78 in log unit) we derived agrees well with that (9.7 in log unit) of \cite{konstantopoulos2010galaxy}, who used archival VLA HI data to map the column density of HI across the system and to derive the masses of all member galaxies. For HCG 59, the angular distance between HCG 59c and HCG 59d is $1^\prime 33^{\prime \prime}$, thus we can take the maximum value of HI flux for HCG 59c and HCG 59d to represent the total flux of them. And then we combined this flux with the HI flux for HCG 59b to derive the total HI flux for HCG 59. For HCG 92, the angular distance between HCG 92b and HCG 92d is $25^{\prime \prime}$. We used the HI mass from integral total line flux in the GBT spectra for HCG 92 in \cite{borthakur2010detection}. For HCG 99, the angular distance between HCG 99a and HCG 99c is $1^\prime 46^{\prime \prime}$. We use the maximum value of the HI flux for these two galaxies as the total HI flux.

Our data selection is based on the availability of HI data. We found, on average, groups with detected HI sources have more spiral galaxy fraction (44.02$\%$ versus 15.74$\%$), more mid-IR active galaxy fraction (42.40$\%$ versus 23.15$\%$) and less mid-IR quiescent galaxy fraction (35.87$\%$ versus 42.59$\%$) compared to groups which do not have observed HI data, and thus it is necessary for us to assess the impact of excluding HI un-detected galaxy groups on our results. To carry out our investigation, we need to estimate the expected HI mass for all the galaxies in our sample.

 We used the formula derived by \cite{haynes1984influence}, which calculated HI mass based on their galaxy morphologies and sizes derived from 324 field galaxies sample:
\begin{equation}
  log(M_{H I, \exp })=a+b \log \left(h D_{25}\right)^{2}-2 \log h
\end{equation}
where $D_{25}$ is the diameter of the galaxy converted to kpc in the B-band at the $25^{th}$ $\ mag\;arcsc^{-2}$, $a$  and $b$ are constants depending on the galaxy’s morphology. $h$ is taken as 0.7 for $H_0$ = 70 $\ km\;s^{-1}$. \cite{healy2021hi} made a detailed summary of previous related work and provided their best estimates of the  values of a and b in their Table 1, thus we use these values in Eq (1). Combined with $D_{25}$ obtained from LEDA, we calculated the predicted HI mass of each galaxy in groups and add them together as each group’s total predicted HI mass. However, the scaling relation in Eq(1) was derived from field galaxies, while galaxies in compact group are known to be deficient in HI mass. Thus we need to make a correction based on the HI deficiency in HCGs.The HI deficiency is defined as $Def_{HI} =log(M(HI)_{predicted}) -log(M(HI)_{observed})$. It is a measure of the deviation of the observed HI mass from the predicted HI mass for a galaxy.
\cite{verdes2001neutral} found that the predicted HI mass for a galaxy based on its optical luminosity and morphology is much larger than the observed HI mass for galaxies in HCGs. The mean value of $Def_{HI}$ is $0.36\pm 0.06$ for their 72 HCG groups sample. To improve the accuracy of the estimated HI mass of our sample, we employ the Monte Carlo (MC) method to estimate the mean  $Def_{HI}$. We produce a random value based on a Gaussian distribution for $M(HI)_{estimated}$ relative errors in each experiment and repeat this process 100,000 times. For the 40 HCG groups with observed HI data, the mean $Def_{HI}$ is $0.28\pm 0.02$. We use this value to correct our estimation of HI mass.
Finally, we derived the corrected HI mass estimates for the total 64 HCG groups, including the 40 groups with observed HI data and the 24 groups with no HI detections. To avoid potential selection effect related to the HI selected subsample, we carried out the entire analysis for both the 40 HCG groups with observed HI data and the total 64 HCG groups. The results are discussed in the following section. The estimated HI mass for HCGs is listed in Table~\ref{tab.B}.

\section{Results}
\label{sect:results}
With the above data, we can study for HCGs the relationship between crossing time and group proprieties. Equation (2) of \cite{hickson1992dynamical} gives a robust estimate of the crossing time for HCGs. $H_{0}{t_{c} }$ is the radio of the crossing time to the approximate age of the universe. Its reciprocal is roughly the maximum number of times a galaxy could have traversed the group since its formation and is thus a measure of the potential dynamical evolution state (\citealt{hickson1992dynamical}). As an important component of galaxies, we also study the HI gas fraction as a function of crossing time. In our investigation, we use the software Matlab to perform the statistical analysis. In particular, we use the function "corrcoef" to study the the correlation, which returns the matrix of correlation coefficients r and the matrix of p, where the p-values are used for testing the hypothesis that there is no relationship between the observed phenomena (null hypothesis). If an off-diagonal element of p is smaller than the significance level (default is 0.05), then the corresponding correlation in R is considered significant.

\subsection{HI mass fraction and crossing time}
\begin{figure}
   \centering
  \includegraphics[width=0.82\textwidth, angle=0]{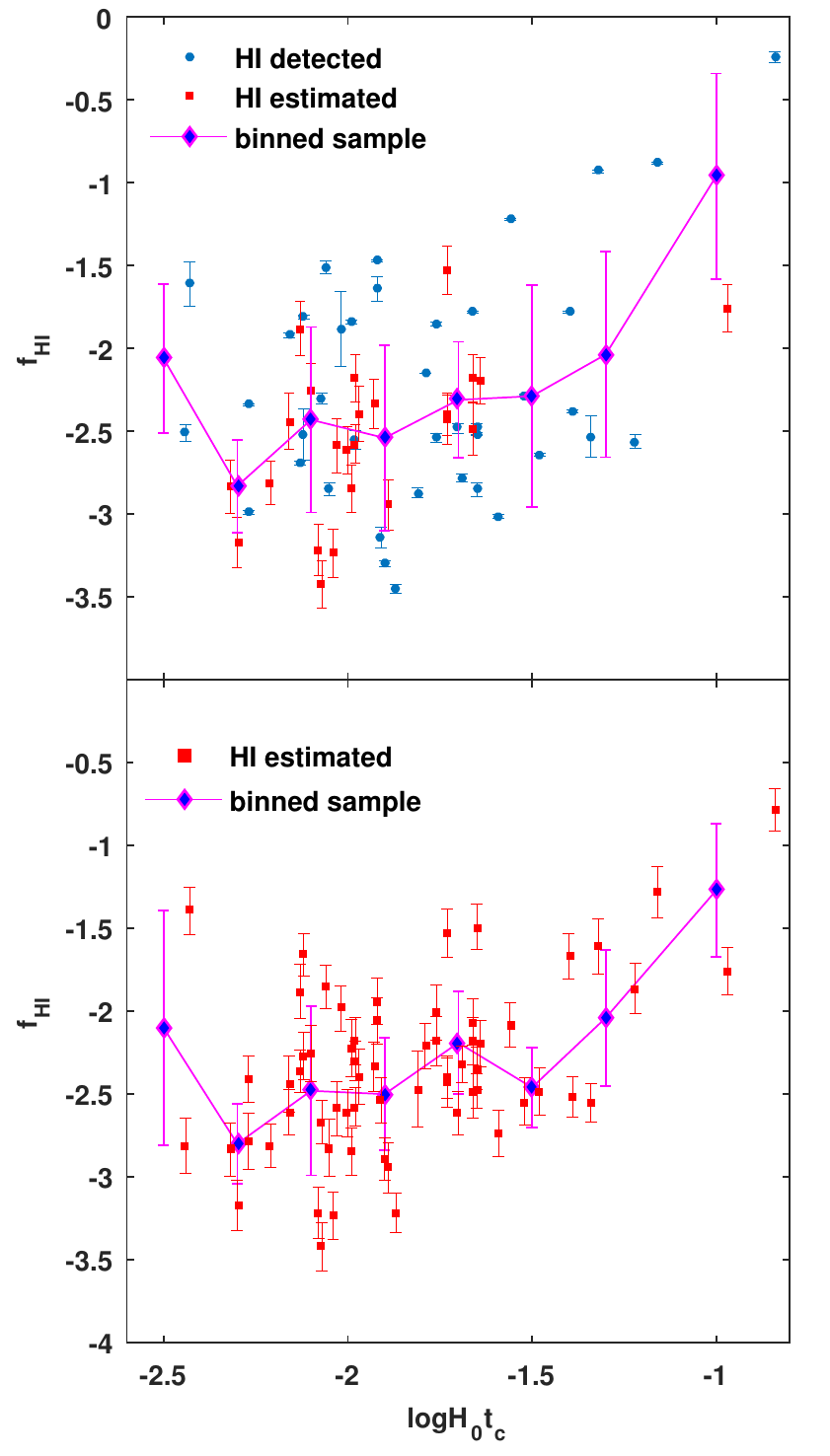}
  \caption{The relationship between the HI gas fraction $f_{HI}$ and the crossing time $H_{0}t_{c}$. In the upper panel, blue circles represent HI detected groups and red squares represent HI estimated groups. The dark blue diamonds are the mean value with a bin size of 0.2 for the total sample. The lower panel shows the results of the estimated HI mass fraction for the total 64 HCG sample.}
  \label{Fig2}
  \end{figure}

In the upper panel in Figure~\ref{Fig2}, we plot the HI mass fraction vs. group crossing time in log unit. We use the group virial mass $ M_{V} $ and HI mass to calculate group HI mass fraction $f_{HI}= \log\left ( M_{HI}/M_{V}   \right )$. The HI detected groups are marked in blue circles and the HI estimated groups are marked in red squares. The Pearson correlation coefficient for the 40 HI detected sample is 0.33 (p=0.04) which indicates the HI mass fraction is weakly correlated with crossing time. However, the corresponding Spearman correlation coefficient is 0.12(p=0.45). After careful examination, we found that the Pearson correlation coefficient is mostly driven by the last data point. To reduce this effect, we include the 24 extra groups with HI mass estimated in the following analysis. We employ the  Monte Carlo (MC) method to estimate the errors in the correlation coefficients.

To conduct MC experiments we need to estimate the error in related parameters. According to the error propagation formula, the relative error of  $H_{0}t_{c}$ and $f_{HI}$is $\sqrt{e_{V }^{2} +e_{R}^{2}}$ and $\sqrt{e_{M_{HI}}^{2}+e_{M_{V}}^{2}}$ respectively, where $e_{V}$ is the relative error of the intrinsic three-dimensional velocity dispersion V, $e_{R}$ is the relative error of the median length of the two-dimensional galaxy-galaxy separation vector R, $e_{M_{HI}}$  is the relative error of HI mass, and $e_{M_{V}}$ is the relative error of viral mass. It is difficlut to estimate the error of viril mass for our HCG groups, while according to \cite{heisler1985estimating}, 75$\%$ of the viral mass estimates lie within $10^{0.25}$  of the correct value for N = 5 groups and within $10^{0.15}$ for N=10. The mean group member number (range from 4 to 7) is 4.6 for our sample, so we use an approximate value of 0.25 for the relative error of the viral mass(in log scale) for the 64 HCG groups. Since the intrinsic three-dimensional velocity dispersion V is proportional to radial velocity dispersion, the relative error of  radial velocity dispersion is the same as V. We conduct Monte Carlo experiments to estimate the relative error in radial velocity dispersion. For each  Monte Carlo realization, we select a random value (assuming Gaussian distribution) within radial velocity error bars for every member galaxy within the group. After repeating this process 1000 times for each group, we get the the relative errors of group radial velocity dispersion for the total 64 HCG groups sample. The error in HI mass is provided in Table~\ref{tab.B}{}. We also calculated the stand deviation in the two-dimensional galaxy-galaxy separation R in each groups. 

\begin{figure}[ht]
  \begin{minipage}[t]{0.495\linewidth}
  \centering
   \includegraphics[width=70mm,height=60mm]{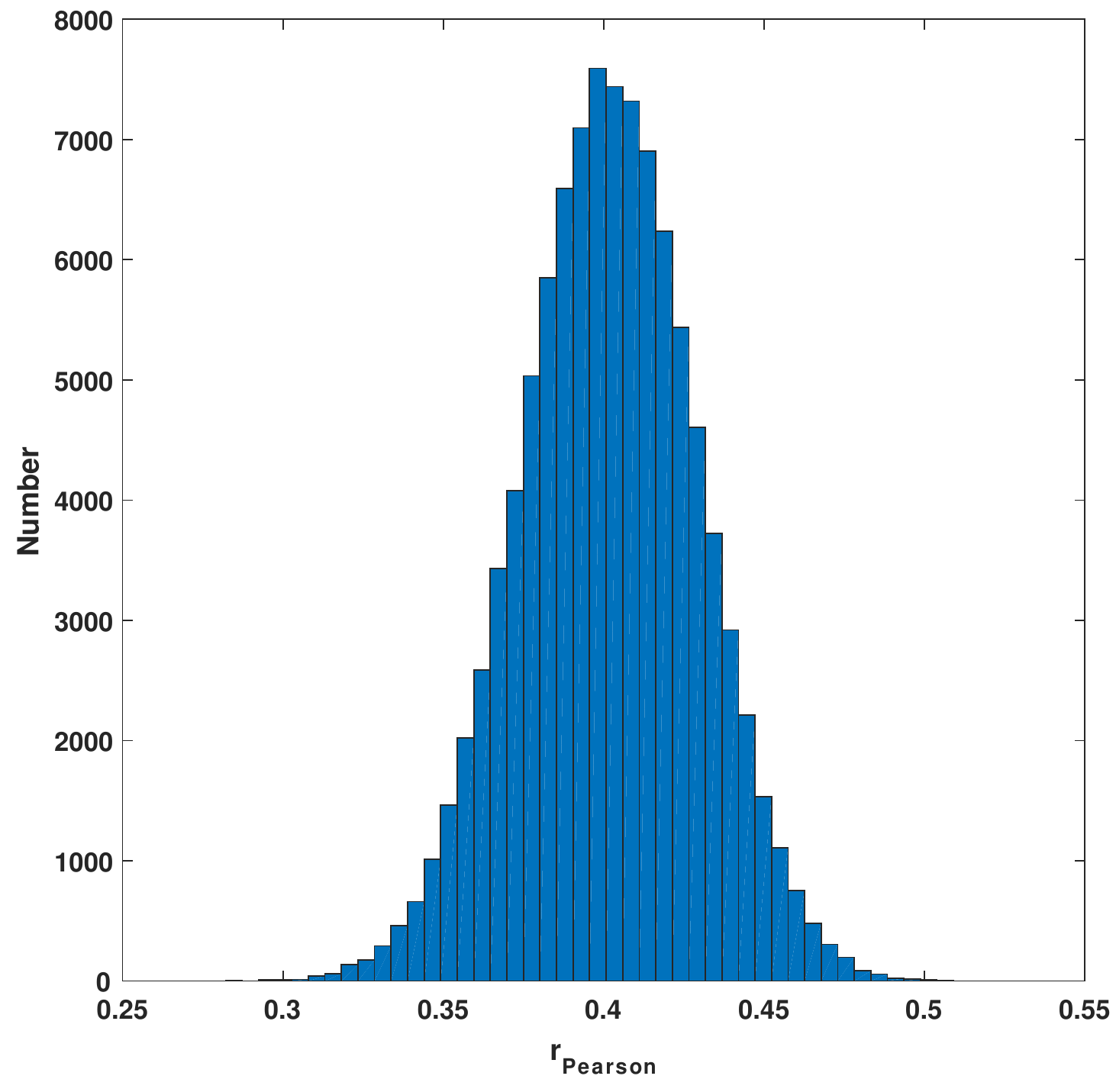}
  \end{minipage}%
  \begin{minipage}[t]{0.495\textwidth}
  \centering
   \includegraphics[width=70mm, height=60mm]{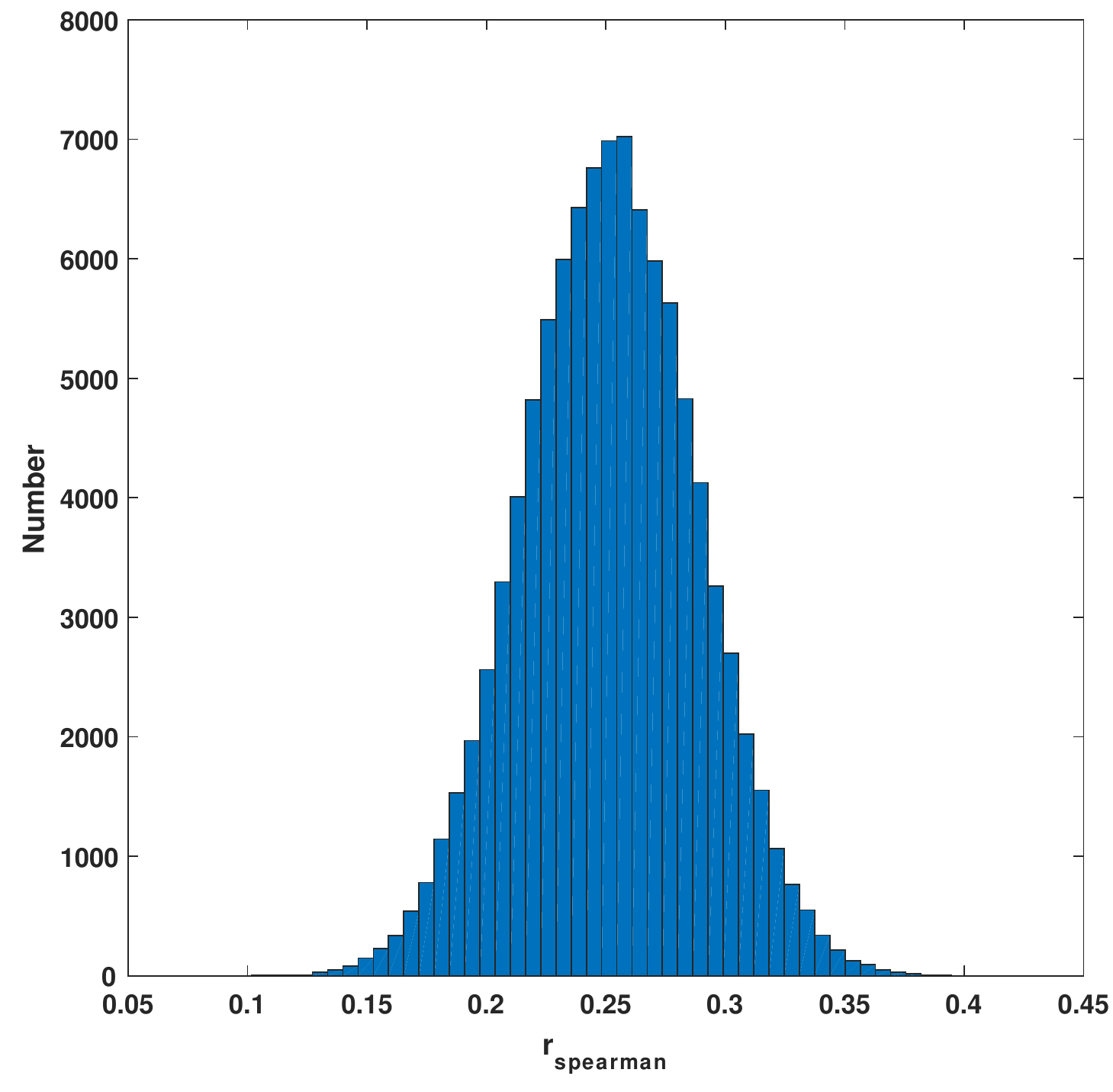}
  \end{minipage}%
  \caption{The Pearson(Left) and Spearman(Right) Correlation Coefficient distributions after 100,000 MC realizations for the 64 HCG groups, bin size=0.005.}
  \label{Fig3}
\end{figure}

We produce a random value based on a Gaussian distribution for  $H_{0}t_{c}$ and $f_{HI}$ relative errors in each experiment and repeat this process 100,000 times. For the 64 galaxy groups sample, the the Pearson correlation coefficient is $r=0.40\pm 0.03$($p=0.001\pm 0.001$) and the Spearman correlation coefficient is $r=0.25\pm 0.04$($p=0.05\pm 0.04$),which means that these two variables are at least weakly correlated. Both correlation coefficients obey the Gaussian distribution shown in Figure~\ref{Fig3}.Finally, we divided the data into several sections with bin size 0.2 in order to reduce the error due to the uncertainty in  $H_{0}t_{c}$. The average value of each bin is marked in dark blue diamonds in the upper panel in Figure~\ref{Fig2} . The error bars are the 1 $\sigma$ standard deviation in each bin and the binned Pearson/Spearman correlation coefficient is 0.68 (p=0.06) and 0.62(p=0.12) respectively. The correlation appears to be stronger in the binned data set.

To compare the trend of estimated HI fraction for the total 64 HCG sample vs. group crossing time with that of groups with HI data, in the lower panel in Figure~\ref{Fig2} we plot the estimated HI fraction vs. group crossing time. Using the same procedure as described above, we derived the correlation coefficients as follows: The Pearson correlation coefficient is $r=0.46\pm 0.03$($p<0.001$) and the Spearman correlation coefficient is $r=0.36\pm 0.04$($p=0.005\pm 0.006$). For the binned data, the Pearson/Spearman correlation coefficient is 0.67 (p=0.07) and 0.60(p=0.13)  respectively. The trend of the total 64 HCG sample with estimated $f_{HI}$ vs. $H_{0}t_{c}$ is similar to that of groups with real HI measurements shown in the upper panel in Figure~\ref{Fig2}.

\subsection{ Galaxy type fraction and crossing time}
As shown in the above section, our investigation on the correlation between HI mass fraction and crossing time suggests that although the relationship found by the 40 HI detected sample shows the same trend as that of the 64 total sample, adding more group data points would increase the reliability of our results. Therefore, we carry on our study based on the 64 total sample in the following sections.

\begin{figure}
   \centering
  \includegraphics[width=\textwidth, angle=0]{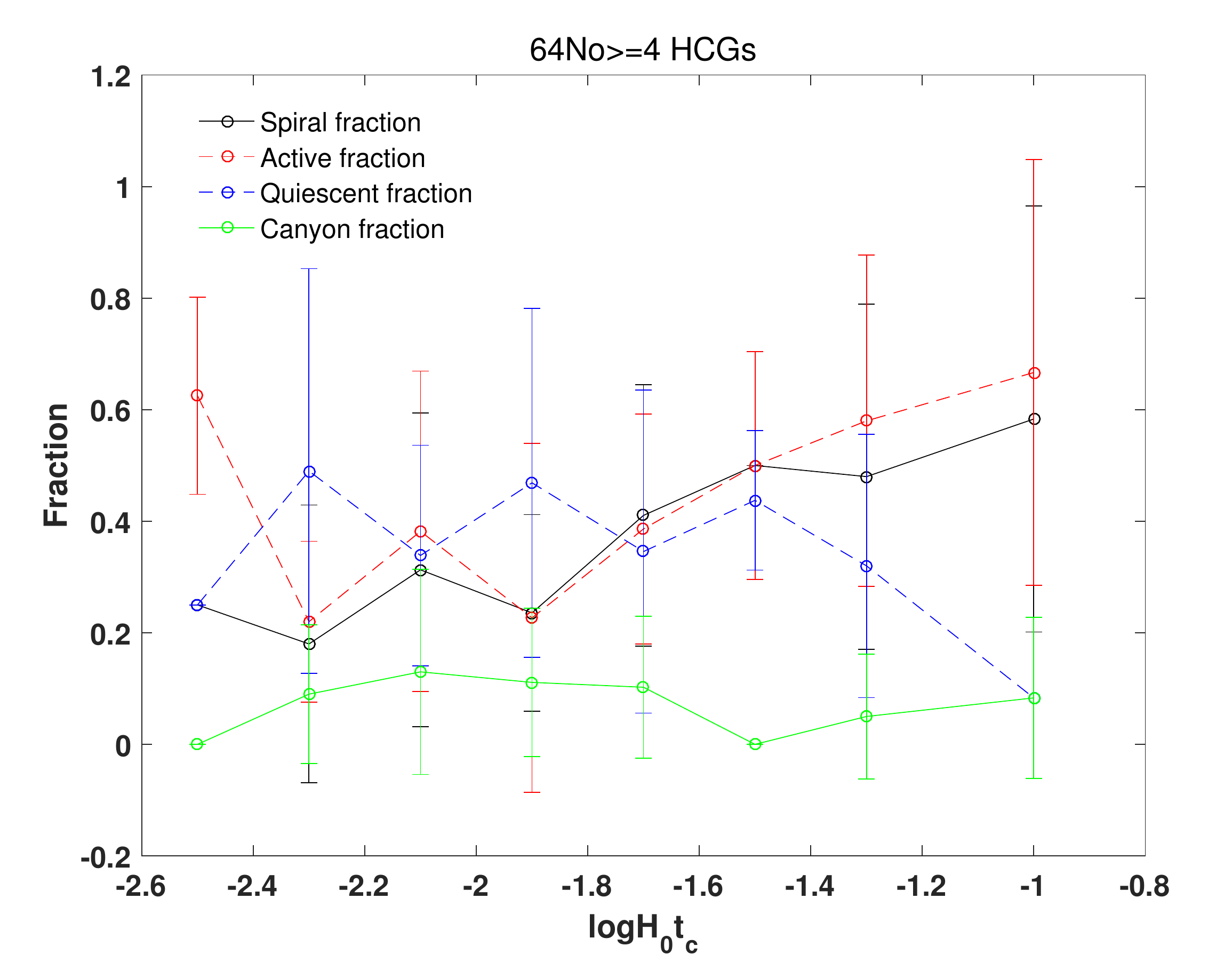}
  \caption{Fraction of spiral, active, quiescent and canyon galaxies plotted versus group crossing time with a bin size 0.2. The objects are color-coded as is shown in upper left corner in the figure. Black line represents spiral fraction, red line represents active fraction, blue line represents quiescent fraction and green line represent canyon fraction.The figure contains data from the total 64 HCG groups. }
  \label{Fig4}
  \end{figure}

We found that the Spearman correlation coefficient of spiral galaxies fraction, active galaxies fraction, quiescent galaxies fraction and canyon galaxies fraction as a function of group crossing time is 0.33, 0.16, -0.09, and -0.07 (p=0.008, 0.20, 0.48, 0.57)respectively. Only the spiral galaxy fraction shows a moderate correlation with group crossing time. The p-values for active galaxies fraction, quiescent galaxies fraction and canyon galaxies fraction are much larger than 0.05, which indicate there is no obvious correlation between these parameters and group crossing time.
To reduce the uncertainty caused by crossing time estimates, we divided the data into several sections with bin size 0.2 which are shown in circles in Figure~\ref{Fig4}. The color-coded circles represent the mean value in each bin with a 1 $\sigma$ standard deviation error bar. The corresponding Spearman correlation coefficient is  0.86, 0.48, -0.33, and -0.16 (p=0.01, 0.24, 0.43, 0.72). The trend is clear that groups with higher crossing times have more spiral galaxies.

\subsection{HI mass and galaxy type fraction}
\begin{figure}[ht]
  \begin{minipage}[t]{0.495\linewidth}
  \centering
   \includegraphics[width=70mm,height=60mm]{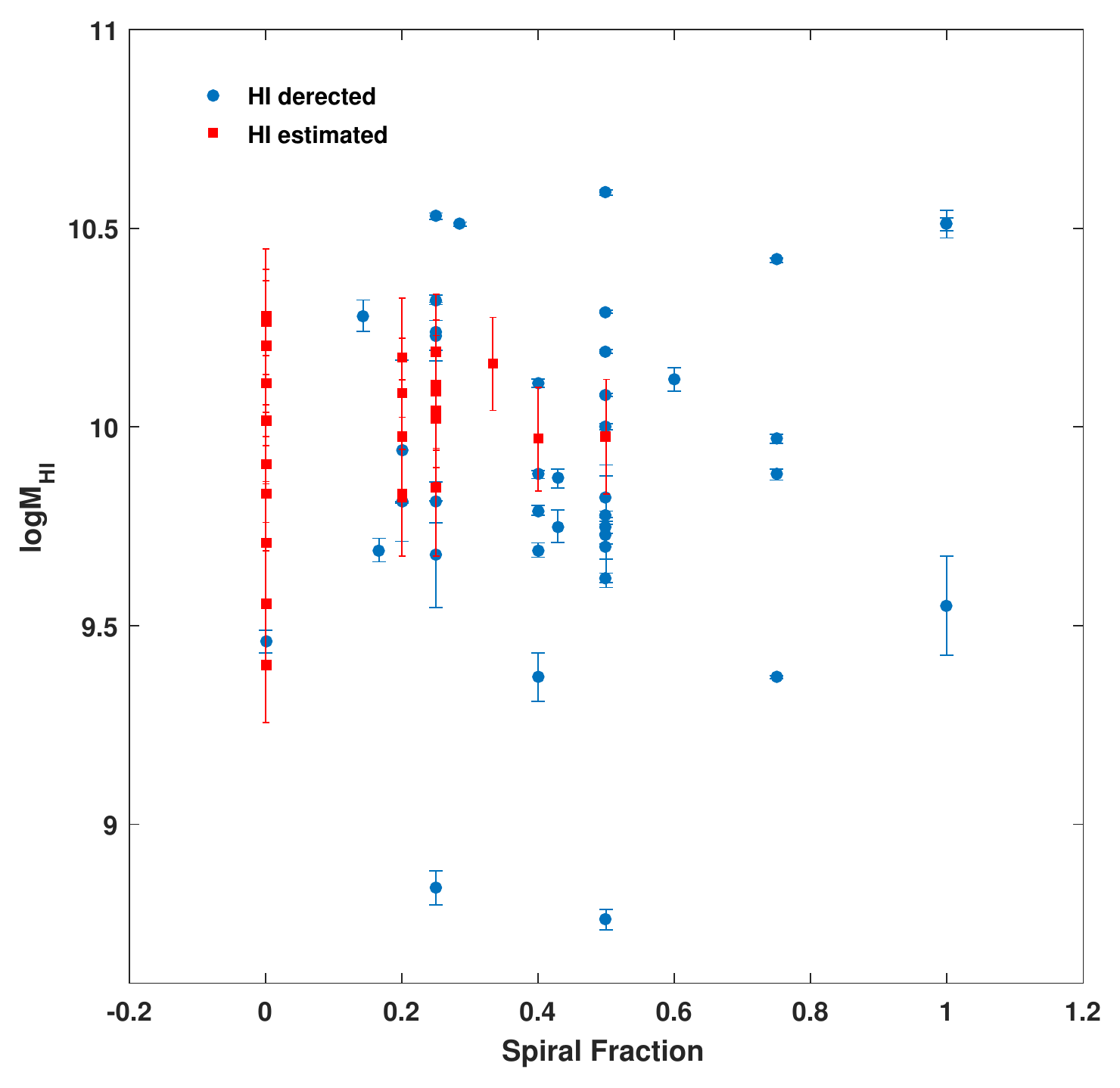}
  \end{minipage}%
  \begin{minipage}[t]{0.495\textwidth}
  \centering
   \includegraphics[width=70mm, height=60mm]{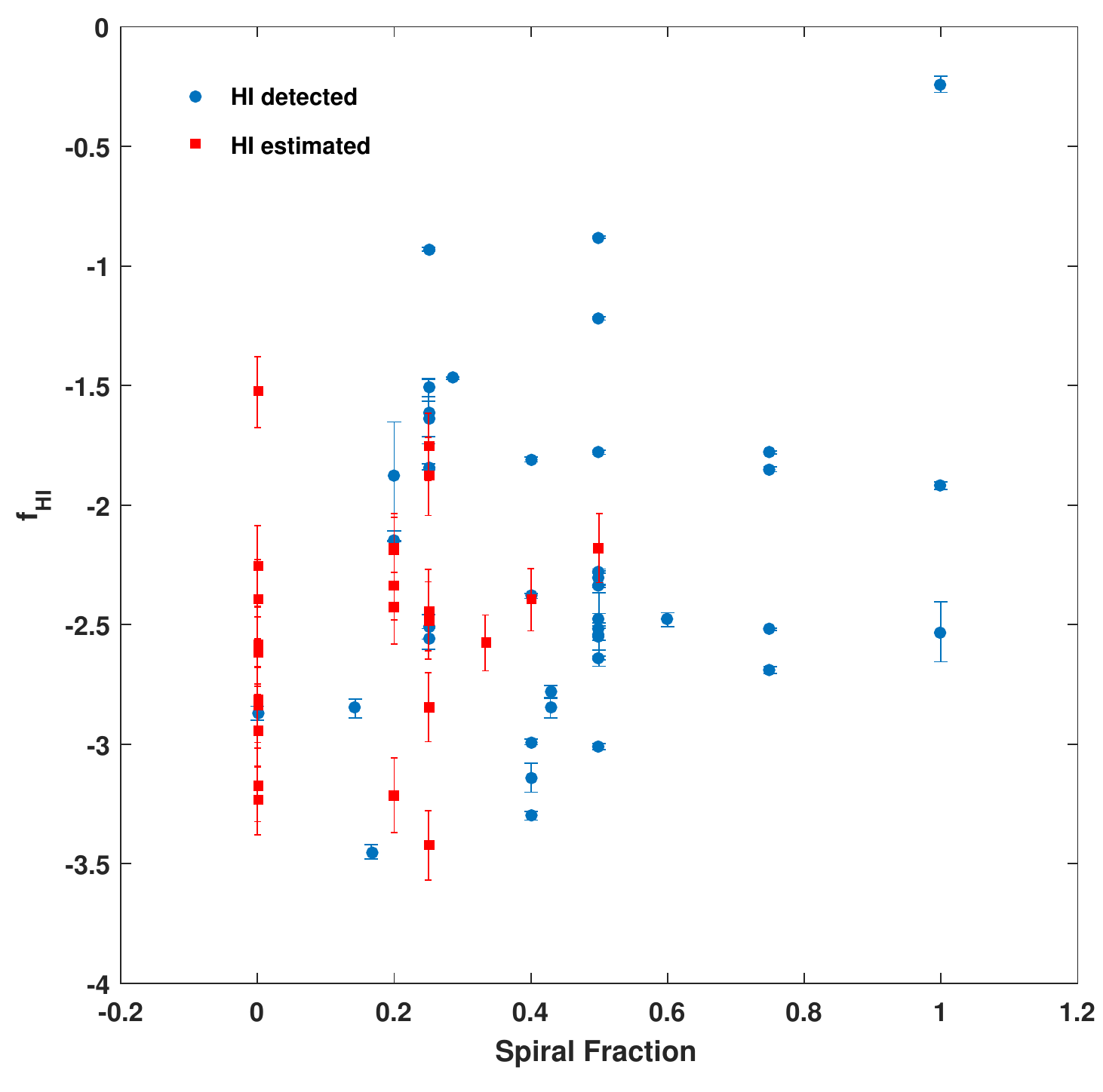}
  \end{minipage}%
  \caption{$\log\left ( M_{HI}  \right ) $ and $\log\left ( f_{HI}  \right ) $ as a function of spiral fraction.Blue circles represent groups with detected HI mass. Red squares represent groups with estimated HI mass.}
  \label{Fig5}
\end{figure}

\begin{figure}[ht]
  \begin{minipage}[t]{0.495\linewidth}
  \centering
   \includegraphics[width=70mm,height=60mm]{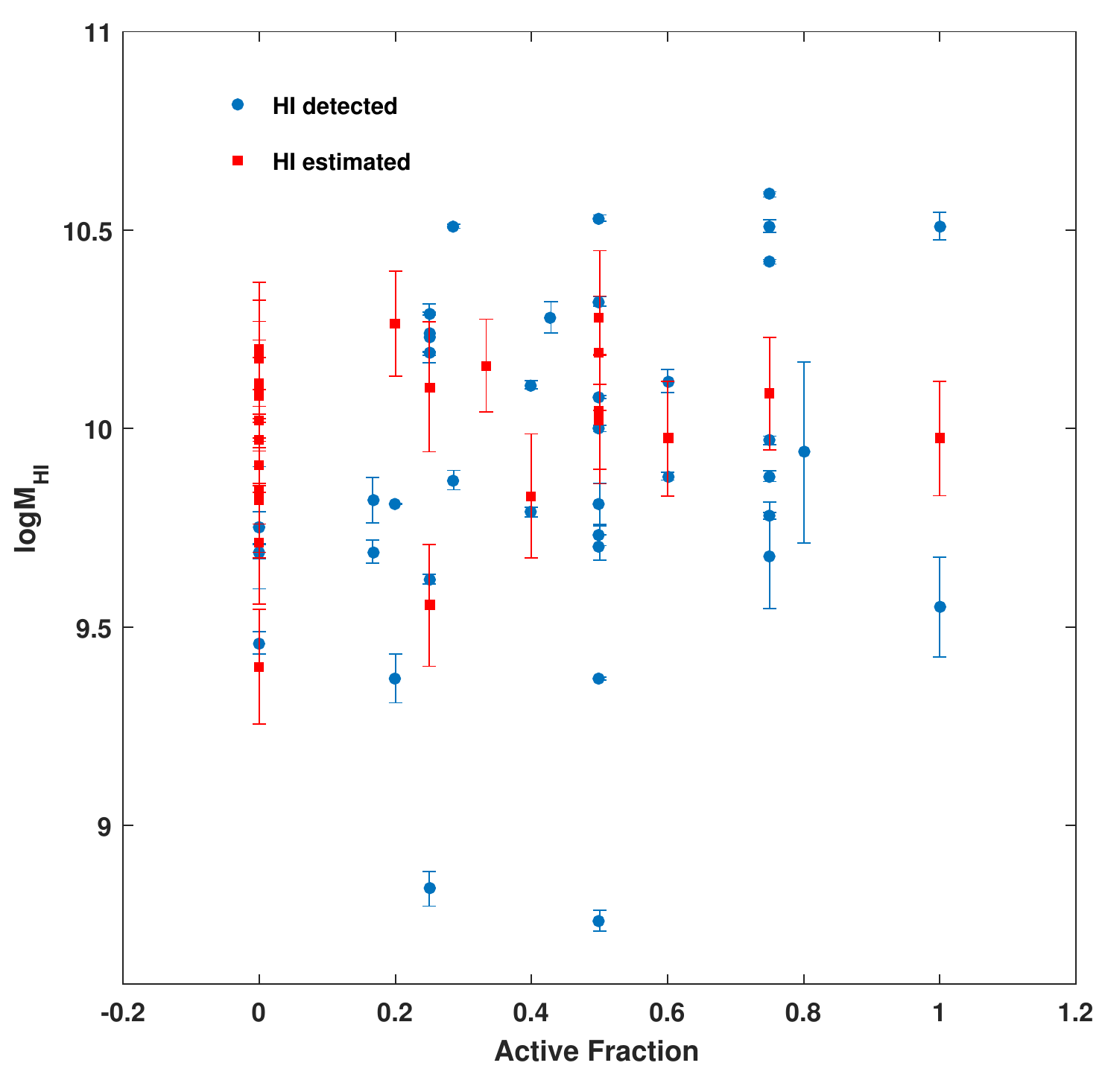}
  \end{minipage}%
  \begin{minipage}[t]{0.495\textwidth}
  \centering
   \includegraphics[width=70mm, height=60mm]{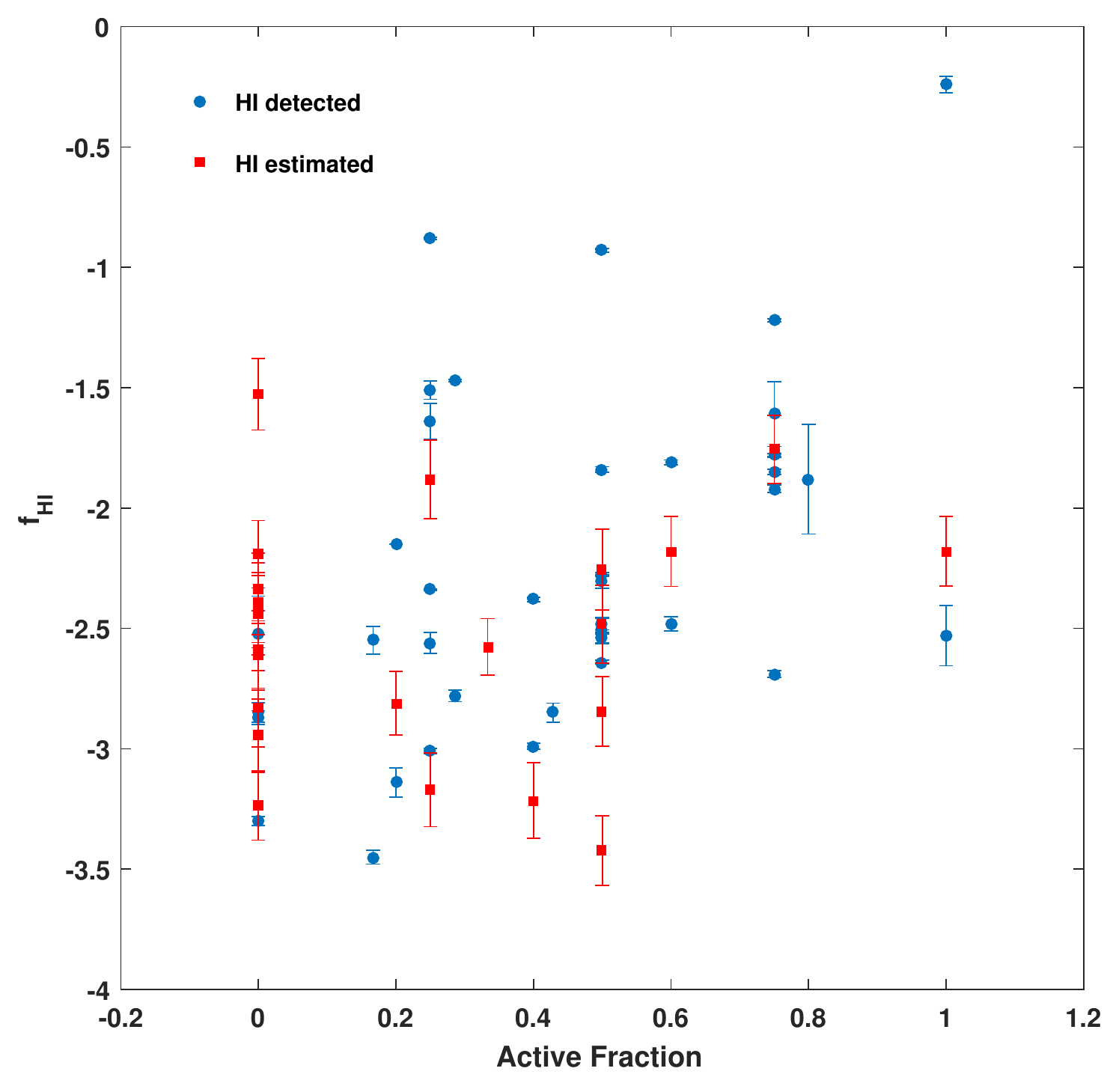}
  \end{minipage}%
  \caption{$\log\left ( M_{HI}  \right ) $ and $\log\left ( f_{HI}  \right ) $ as a function of active fraction.Blue circles represent groups with detected HI mass. Red squares represent groups with estimated HI mass.}
  \label{Fig6}
\end{figure}

Spiral galaxies are usually gas-rich, therefore we expect the HI mass or HI mass fraction would increase for groups with more spiral members. Figure~\ref{Fig5} shows HI mass (Left) and HI mass fraction (Right) vs. Spiral fraction. The HI detected groups are marked in blue circles and the HI estimated groups are marked in red squares. The Spearman correlation coefficient is 0.01 (p=0.94) and 0.24 (p=0.06) respectively for the total 64 HCG groups, which indicates there is a weak correlation between the spiral fraction and the HI mass fraction. Furthermore, we check the HI mass/HI mass fraction as a function of mid-IR active (meaning  actively star-forming ) fraction shown in Figure~\ref{Fig6}. The Spearman correlation coefficient is 0.21 (p=0.09) and 0.40 (p=0.001) respectively, which indicates that the actively star-forming galaxy fraction is weakly correlated with HI mass and moderately correlated with HI mass fraction. 

Our investigation on these relationships based on estimated HI mass fraction for the total 64 HCG sample including both the HI detected and the HI undetected groups shows similar results. For the 40 HI detected sample the results are mostly consistent with that of the total 64 HCG sample with one exception. The Spearman correlation coefficient of $f_{HI}$ vs. $f_{spiral}$ is 0.12(p=0.47) for the 40 group sample, which indicates there is no obvious correlation between the HI mass fraction and the spiral fraction. In other words, the weak correlation between the spiral fraction and HI mass fraction for the total 64 sample would not exist if we excluded the 24 HI estimated groups. Note that most of the 24 HI estimated groups(marked in red squares) have small spiral fraction and low HI mass fraction, as is shown in the right figure in Fig~\ref{Fig5}. The exclusion of groups at the small spiral fraction end could produce strong bias when we investigate the correlation between $f_{HI}$ and $f_{spiral}$. To reduce this influence, in the future, we plan to collect more HI data from the FAST extragalactic HI sky survey(19 of the 24 HI undetected groups are within FAST visible area) and extend our CG sample to improve the statistics.

%
 
%


\section{Discussion}
\label{sect:discussion}

\subsection{Group crossing time}
Note that the groups extracted from SDSS group catalog in \cite{ai2018evolution} are normal groups with more ($N\ge 8$) galaxy members compared to HCGs. The crossing time $H_{0}{t_{c} }$  in our compact group sample is smaller (ranges from -2.4 to -0.8 in log unit, mean value -1.79) than that in the SDSS groups (ranges from -1.6 to -0.6 in log unit), indicating that compact groups are more dynamically evolved than SDSS groups. This is not surprising, as high-density environment of CGs has accelerated the evolution of galaxies. The virial mass of of our sample ranges from 10.75 to 13.14 with an average value 12.14, while the virial mass of group sample in \cite{ai2018evolution} ranges from 13 to 14.5. The HI mass fraction of CGs in our sample, ranging from-3.45 to -0.24 with a mean value -2.2, is higher than the HI mass fraction of SDSS group sample in \cite{ai2018evolution}, which ranges from -4.1 to -2.1. This is in consistent with the study by \cite{guo2020direct}, who found the gas fraction ($M_{HI}/M_{h}  $ ) decreases when the halo mass increases. They measured the total HI mass in halos by stacking HI spectra of entire groups within the ALFALFA survey using SDSS DR7 group catalog. Despite the fact that HCGs have smaller $H_{0}{t_{c} }$ but higher $f_{HI}$ , the result of our test with HI mass fraction and crossing time is consistent with that of \cite{ai2018evolution} e.g., groups with larger crossing time tend to be HI rich and groups with smaller crossing time tend to be HI poor. 

In the lower panel in Figure~\ref{Fig2}, we plot the estimated $f_{HI}$ vs.  $H_{0}{t_{c} }$ for the total 64 HCG sample. This result shows similar trend. Since our estimated HI mass is derived from $D_{25}$, the correlation between estimated $f_{HI}$ and $H_{0}{t_{c} }$ indicates that groups with larger crossing time tend to have galaxy members with larger B-band major diameter. Meanwhile, according to \cite{haynes1984influence}, $log D_{25}$ is proportional to $log L_{B}$, where optical luminosity $L_{B}$ is defined by Eq (8) in their paper. Our result suggests that younger groups with larger crossing times also have more blue band luminous galaxy members.

One caveat of our results is that the level of HI deficiency varies widely in our HCGs sample, thus our correction for the HI deficiency in HCGs using a single value can have systematic uncertainties that can affect some of the correlations with crossing time. A more accurate method for estimating the HI mass of galaxies in compact groups is needed in future studies. Another drawback with our results is the lack of data points with large crossing time in the range of (-1.2~-0.8). Such limitation reduces the statistical significance of the results. A larger sample of HCG galaxies with observed HI data from future FAST HI galaxy surveys is needed to improve the statistical results.

 \subsection{Galaxy types}
The positive correlation between spiral galaxy fraction and group crossing time obtained from sub-sample of HCGs is consistent with the result of \cite{hickson1992dynamical}, which is in line with our expectation since with smaller crossing time, galaxy merging would convert spiral galaxies into elliptical galaxies. However, there is no correlation between active galaxy fraction and crossing time in our HCG subsample. This may be related to new galaxy members being accreted into the group. As suggested by \cite{durbala2007seyfert}, compact groups form by slow sequential acquisition of neighbors from surrounding larger scale structure. New intruders are usually late-type (Sbc and later) spirals, still rich in gas. When they fall into the group, these galaxies increase their star formation and quickly lose most of their ISM and their morphology transform into spiral bulges or into early-type (E-S0) galaxies. Active nuclei could also be stimulated by the residual unstripped gas. In the 64-group sample, only 67 (65$\%$) actively star-forming galaxies are spirals while the rest active galaxies are  early-type (E-S0) galaxies, suggesting that the time scale of active star formation is longer than morphology change.
In our HCG sample, the canyon galaxy fraction remains low and shows little changes with crossing time, indicating that the transition of galaxies from actively star forming to quiescent is a fast process, as has been confirmed by many previous work based on UV and IR data (\citealt{johnson2007infrared,tzanavaris2010ultraviolet+,walker2010mid,bitsakis2011mid}). On the other hand, the star forming phase can last for a long time in HCGs, thus we see no correlation between the active galaxy fraction and crossing time.

\subsection{ HI content}
As shown in Fig~\ref{Fig5} and Fig~\ref{Fig6}, both the spiral galaxy fraction and active galaxy fraction shows at least weak correlation with HI mass fraction. These results make sense, as spiral galaxies are usually gas rich and groups with higher HI mass fraction have more raw fuel for star-forming. A similar result was found by \cite{walker2016global} for a HCG $\&$ RSCG combined catalog that HI mass fraction increases with increasing spiral fraction with large scatter and galaxies in HI-rich groups tend to be actively star-forming, while galaxies in HI-poor groups tend to be mid-IR quiescent.

Another fact is that the average two-dimensional galaxy-galaxy angular distance is $2^\prime 46^{\prime \prime}$ in our 40 HCG groups sample, and the HI mass of 25(62.5$\%$) HCG groups are derived from integral total HI line flux including all group members in the single-dish telescope spectra (Effelsberg,Green Bank 91m, Green Bank 42m and Nancy). Our census on group HI mass are likely to include both HI in the galaxies and in the intra-group medium. As suggested by \cite{borthakur2010detection}, who compared new Green Bank Telescope (GBT) 21 cm HI spectra with corresponding VLA spectra for a complete distance-limited sample of 22 HCGs and found that there exists significant HI excess in the GBT spectra, varying from 5$\%$-81$\%$ with and an average of 36$\%$. They also found that the excess gas increases with the evolutionary stage of the group. To get a further investigation of how this effect would affect our results, higher resolution interferometer data are needed to study the HI mass distribution in these groups.

\section{Conclusions}
\label{sect:conclusion}
Our main conclusions are summarized as follows: 

1. We selected a subsample containing 64 galaxy groups from HCG catalog using n $\ge$ 4 as data selection criteria. We obtained group morphology data from the LEDA and the IR classification from \cite{zucker2016hierarchical}. 40 of 64 groups have HI observed fluxes from $\alpha.100$ catalog, Springob/Cornell HI data catalog and other direct measurements of HCGs in the literature. To reduce the HI selection effect, we used the scaling relation to estimate the HI mass for the other 24 groups with no HI detection. 

2. Our analysis with both the 40 HI detected sample and the 64 total sample shows a weak correlation between the HI mass fraction and the group crossing time, similar to the results found in SDSS groups in \cite{ai2018evolution}. This result suggests that as galaxies in CGs evolves, the HI mass fraction decreases with dynamically evolutionary stages. 

3. In the n $\ge$ 4 subsample, we found that group spiral fraction and group crossing time shows the same correlation as that in \cite{hickson1992dynamical}. We found there is no obvious correlation between actively star-forming galaxy fraction, quiescent galaxy fraction, canyon galaxy fraction (classified by their mid-IR color in \cite{zucker2016hierarchical}) and group crossing time.

4. We found that the HI content is correlated with the spiral fraction and active galaxy fraction in HCGs. These results are consistent with those reported by \cite{walker2016global} that galaxies in HI-rich groups tend to be actively star-forming, while galaxies in HI-poor groups tend to be mid-IR quiescent.

\begin{acknowledgements}
We thank the anonymous referee for his/her comments and constructive suggestions. This work is supported by the National Key R$\&$D Program of China No.2107YFA0402600. We acknowledge the work of the entire ALFALFA collaboration team in observing, flagging, and extracting the catalog of galaxies used in this work. We acknowledge the usage of the Hyper-Leda database (http://leda.univ-lyon1.fr) and of the Extragalactic Distance Database (EDD) (http://edd.ifa.hawaii.edu). TOPCAT is an interactive graphical tool for analysis and manipulation of tabular data (http://www.star.bris.ac.uk/~mbt/topcat/).
\end{acknowledgements}

\appendix                  

\section{Table of group properties}
The column heading for this table are (1) Group number in HCGs; (2) group redshift; (3) number of galaxies in group; (4) estimated intrinsic three-dimensional velocity dispersion; (5) relative error of V; (6) median projected separation; (7) relative error of R; (8) crossing time; (9) virial mass; (10) Group HI mass; (11) Group HI mass error; (12) Estimated group HI mass; (13) relative error of estimated group HI mass; (14) source of data: 1,$\alpha.100$; 2,EDD; 3,\cite{borthakur2010detection}; 4,\cite{jones2019evolution}; 6, \cite{huchtmeier1997hi}; 7, corrected estimation;(15) spiral galaxy number; (16) Active galaxy number; (17) Quiescent galaxy number; (18) Canyon galaxy number.
\begin{landscape}
\begin{longtable}[c]{cccccccccccccccccc}
\caption{The resulting values for the HCGs}
\label{tab.B}\\
\hline
GroupID & Z & n & log(V) & slog(V) & log(R){[}kpc{]} & slog(R) & log(Ho.tc) & logMv & HI & HI\_err & HI\_est & logHI\_er & Ref. & S & A & Q & C \\ \hline
\endfirsthead
\multicolumn{18}{c}%
{{\bfseries Table \thetable\ continued from previous page}} \\
\hline
GroupID & Z & n & log(V) & slog(V) & log(R){[}kpc{]} & slog(R) & log(Ho.tc) & logMv & HI & HI\_err & HI\_est & logHI\_er & Ref. & S & A & Q & C \\ \hline
\endhead
\hline
\endfoot
\endlastfoot
1 & 0.0339 & 4 & 2.12 & 0.054 & 1.69 & 0.21 & -1.32 & 11.46 & 10.53 & 8.81 & 10.13 & 0.17 & 1 & 1 & 2 & 1 & 1 \\
6 & 0.0379 & 4 & 2.63 & 0.029 & 1.4 & 0.20 & -2.12 & 12.27 & 9.75 & 9.30 & 10.27 & 0.14 & 6 & 2 & 0 & 2 & 1 \\
7 & 0.0141 & 4 & 2.16 & 0.050 & 1.66 & 0.19 & -1.4 & 11.56 & 9.78 & 8.06 & 10.17 & 0.14 & 1,2 & 2 & 3 & 1 & 0 \\
8 & 0.0545 & 4 & 2.89 & 0.014 & 1.46 & 0.13 & -2.32 & 12.945 & 10.11 & 9.95 & 10.39 & 0.16 & 7 & 0 & 0 & 3 & 1 \\
10 & 0.0161 & 4 & 2.56 & 0.016 & 1.97 & 0.22 & -1.48 & 12.64 & 10.00 & 8.25 & 10.43 & 0.14 & 1 & 2 & 2 & 2 & 0 \\
12 & 0.0485 & 5 & 2.62 & 0.022 & 1.77 & 0.17 & -1.73 & 12.605 & 10.17 & 9.99 & 10.45 & 0.15 & 7 & 1 & 0 & 1 & 0 \\
13 & 0.0411 & 5 & 2.44 & 0.050 & 1.67 & 0.18 & -1.66 & 12.155 & 9.97 & 9.77 & 10.25 & 0.14 & 7 & 1 & 3 & 1 & 0 \\
15 & 0.0228 & 6 & 2.86 & 0.015 & 1.89 & 0.19 & -1.87 & 13.14 & 9.69 & 8.52 & 10.20 & 0.12 & 1 & 1 & 1 & 4 & 0 \\
16 & 0.0132 & 4 & 2.31 & 0.038 & 1.65 & 0.20 & -1.56 & 11.81 & 10.59 & 8.76 & 10.00 & 0.13 & 4 & 2 & 3 & 1 & 0 \\
17 & 0.0603 & 5 & 2.18 & 0.066 & 1.35 & 0.21 & -1.73 & 11.435 & 9.91 & 9.72 & 10.18 & 0.15 & 7 & 0 & 0 & 1 & 1 \\
23 & 0.0161 & 4 & 2.44 & 0.050 & 1.82 & 0.12 & -1.52 & 12.36 & 10.08 & 8.01 & 10.09 & 0.14 & 3 & 2 & 2 & 2 & 0 \\
24 & 0.0305 & 5 & 2.51 & 0.037 & 1.47 & 0.19 & -1.93 & 12.155 & 9.82 & 9.62 & 10.10 & 0.15 & 7 & 1 & 0 & 2 & 1 \\
25 & 0.0212 & 4 & 2.13 & 0.068 & 1.68 & 0.22 & -1.16 & 11.07 & 10.19 & 8.24 & 10.07 & 0.15 & 3 & 2 & 1 & 1 & 0 \\
26 & 0.0316 & 7 & 2.52 & 0.027 & 1.5 & 0.22 & -1.92 & 11.98 & 10.51 & 8.60 & 10.32 & 0.14 & 3 & 2 & 2 & 0 & 0 \\
27 & 0.0874 & 4 & 2.1 & 0.090 & 2.03 & 0.20 & -0.97 & 11.845 & 10.09 & 9.88 & 10.36 & 0.14 & 7 & 1 & 3 & 0 & 1 \\
30 & 0.0154 & 4 & 2.04 & 0.059 & 1.71 & 0.12 & -1.22 & 11.4 & 8.84 & 7.84 & 9.81 & 0.15 & 3 & 1 & 1 & 2 & 0 \\
32 & 0.0408 & 4 & 2.55 & 0.027 & 1.79 & 0.21 & -1.66 & 12.505 & 10.02 & 9.87 & 10.30 & 0.16 & 7 & 1 & 2 & 2 & 0 \\
33 & 0.026 & 4 & 2.42 & 0.026 & 1.39 & 0.20 & -1.92 & 11.88 & 10.24 & 9.47 & 10.10 & 0.14 & 5 & 1 & 1 & 2 & 0 \\
34 & 0.0307 & 4 & 2.74 & 0.020 & 1.19 & 0.20 & -2.44 & 12.32 & 9.81 & 8.88 & 9.79 & 0.17 & 2 & 1 & 2 & 1 & 0 \\
35 & 0.0542 & 6 & 2.74 & 0.015 & 1.65 & 0.20 & -1.98 & 12.735 & 10.16 & 9.86 & 10.43 & 0.12 & 7 & 2 & 2 & 3 & 0 \\
37 & 0.0223 & 5 & 2.84 & 0.012 & 1.46 & 0.19 & -2.27 & 12.78 & 9.79 & 8.24 & 10.27 & 0.17 & 3 & 2 & 2 & 1 & 1 \\
39 & 0.0701 & 4 & 2.52 & 0.026 & 1.44 & 0.19 & -1.98 & 12.155 & 9.97 & 9.77 & 10.25 & 0.14 & 7 & 2 & 4 & 0 & 0 \\
40 & 0.0223 & 5 & 2.4 & 0.018 & 1.18 & 0.19 & -2.12 & 11.69 & 9.88 & 8.24 & 10.31 & 0.13 & 3 & 2 & 3 & 2 & 0 \\
42 & 0.0133 & 4 & 2.56 & 0.019 & 1.65 & 0.20 & -1.81 & 12.33 & 9.46 & 8.28 & 10.14 & 0.23 & 6 & 0 & 0 & 2 & 0 \\
43 & 0.033 & 5 & 2.58 & 0.023 & 1.77 & 0.17 & -1.7 & 12.6 & 10.12 & 8.95 & 10.26 & 0.13 & 1 & 3 & 3 & 1 & 1 \\
44 & 0.0046 & 4 & 2.34 & 0.066 & 1.58 & 0.17 & -1.65 & 11.89 & 9.37 & 7.32 & 9.82 & 0.13 & 1 & 3 & 2 & 1 & 1 \\
46 & 0.027 & 4 & 2.75 & 0.013 & 1.6 & 0.22 & -2.04 & 12.635 & 9.40 & 9.20 & 9.67 & 0.14 & 7 & 0 & 0 & 1 & 2 \\
50 & 0.1392 & 5 & 2.91 & 0.018 & 1.59 & 0.18 & -2.21 & 13.075 & 10.26 & 10.02 & 10.54 & 0.13 & 7 & 0 & 1 & 0 & 0 \\
51 & 0.0258 & 5 & 2.61 & 0.025 & 1.77 & 0.24 & -1.73 & 12.365 & 9.97 & 9.72 & 10.24 & 0.13 & 7 & 2 & 0 & 3 & 2 \\
55 & 0.0526 & 4 & 2.52 & 0.058 & 1.28 & 0.18 & -2.13 & 11.985 & 10.10 & 9.96 & 10.38 & 0.16 & 7 & 1 & 1 & 2 & 0 \\
56 & 0.027 & 5 & 2.45 & 0.034 & 1.33 & 0.24 & -2.02 & 11.82 & 9.94 & 9.66 & 10.11 & 0.13 & 5 & 1 & 4 & 1 & 0 \\
57 & 0.0304 & 7 & 2.66 & 0.030 & 1.86 & 0.22 & -1.69 & 12.65 & 9.87 & 8.62 & 10.61 & 0.11 & 1 & 3 & 2 & 3 & 1 \\
58 & 0.0207 & 5 & 2.44 & 0.015 & 1.95 & 0.16 & -1.39 & 12.49 & 10.11 & 8.48 & 10.25 & 0.12 & 1 & 2 & 2 & 3 & 0 \\
59 & 0.0135 & 4 & 2.51 & 0.027 & 1.33 & 0.17 & -2.07 & 12 & 9.70 & 8.57 & 9.61 & 0.13 & 1,2 & 2 & 2 & 0 & 1 \\
60 & 0.0625 & 4 & 2.85 & 0.043 & 1.75 & 0.27 & -1.99 & 13.035 & 10.19 & 9.99 & 10.46 & 0.14 & 7 & 1 & 2 & 1 & 1 \\
62 & 0.0137 & 4 & 2.69 & 0.017 & 1.43 & 0.25 & -2.16 & 12.285 & 9.85 & 9.71 & 10.12 & 0.17 & 7 & 1 & 0 & 3 & 1 \\
65 & 0.0475 & 5 & 2.74 & 0.017 & 1.66 & 0.22 & -1.97 & 12.595 & 10.20 & 10.06 & 10.48 & 0.17 & 7 & 0 & 0 & 4 & 1 \\
66 & 0.0699 & 4 & 2.72 & 0.025 & 1.51 & 0.21 & -2.1 & 12.535 & 10.28 & 10.14 & 10.55 & 0.17 & 7 & 0 & 2 & 1 & 0 \\
67 & 0.0245 & 4 & 2.56 & 0.026 & 1.69 & 0.24 & -1.76 & 12.27 & 9.73 & 8.49 & 10.37 & 0.15 & 3 & 2 & 2 & 2 & 0 \\
68 & 0.008 & 5 & 2.42 & 0.019 & 1.52 & 0.19 & -1.79 & 11.96 & 9.81 & 7.36 & 10.02 & 0.14 & 3 & 1 & 1 & 4 & 0 \\
69 & 0.0294 & 4 & 2.58 & 0.029 & 1.48 & 0.21 & -1.99 & 12.16 & 10.32 & 8.76 & 10.21 & 0.17 & 1 & 1 & 2 & 1 & 1 \\
70 & 0.0636 & 4 & 2.31 & 0.080 & 1.86 & 0.17 & -1.34 & 12.08 & 9.55 & 9.01 & 9.80 & 0.12 & 5 & 4 & 4 & 0 & 0 \\
72 & 0.0421 & 4 & 2.66 & 0.018 & 1.55 & 0.21 & -2 & 12.445 & 9.83 & 9.63 & 10.11 & 0.14 & 7 & 0 & 0 & 4 & 0 \\
74 & 0.0399 & 5 & 2.73 & 0.024 & 1.59 & 0.22 & -2.03 & 12.605 & 10.02 & 9.86 & 10.29 & 0.16 & 7 & 0 & 0 & 2 & 1 \\
75 & 0.0416 & 6 & 2.66 & 0.031 & 1.57 & 0.24 & -1.98 & 12.37 & 9.82 & 8.94 & 10.35 & 0.12 & 2 & 3 & 1 & 1 & 1 \\
76 & 0.034 & 7 & 2.62 & 0.019 & 1.86 & 0.19 & -1.65 & 12.6 & 9.75 & 8.72 & 10.40 & 0.11 & 1 & 3 & 0 & 5 & 1 \\
79 & 0.0145 & 4 & 2.36 & 0.042 & 0.83 & 0.16 & -2.43 & 11.29 & 9.68 & 9.17 & 10.17 & 0.14 & 3 & 1 & 3 & 1 & 0 \\
80 & 0.031 & 4 & 2.67 & 0.020 & 1.4 & 0.19 & -2.16 & 12.43 & 10.51 & 9.07 & 10.10 & 0.14 & 6 & 4 & 3 & 1 & 0 \\
81 & 0.0499 & 4 & 2.43 & 0.046 & 1.26 & 0.19 & -2.06 & 11.74 & 10.23 & 9.17 & 10.16 & 0.13 & 1 & 1 & 1 & 0 & 2 \\
82 & 0.0362 & 4 & 3.03 & 0.008 & 1.85 & 0.16 & -2.07 & 13.465 & 10.04 & 9.84 & 10.32 & 0.14 & 7 & 1 & 2 & 2 & 0 \\
83 & 0.0531 & 5 & 2.89 & 0.028 & 1.7 & 0.16 & -2.08 & 13.045 & 9.83 & 9.66 & 10.11 & 0.16 & 7 & 1 & 2 & 2 & 0 \\
84 & 0.0556 & 5 & 2.52 & 0.042 & 1.77 & 0.24 & -1.64 & 12.275 & 10.08 & 9.87 & 10.36 & 0.14 & 7 & 1 & 0 & 1 & 1 \\
85 & 0.0393 & 4 & 2.8 & 0.014 & 1.39 & 0.19 & -2.3 & 12.725 & 9.55 & 9.38 & 9.83 & 0.15 & 7 & 0 & 1 & 3 & 0 \\
86 & 0.0199 & 4 & 2.66 & 0.020 & 1.67 & 0.12 & -1.89 & 12.655 & 9.71 & 9.53 & 9.99 & 0.15 & 7 & 0 & 0 & 4 & 0 \\
89 & 0.0297 & 4 & 1.72 & 0.097 & 1.77 & 0.22 & -0.84 & 10.75 & 10.51 & 9.41 & 10.24 & 0.13 & 5 & 4 & 4 & 0 & 0 \\
90 & 0.0088 & 4 & 2.22 & 0.035 & 1.47 & 0.31 & -1.65 & 11.24 & 8.76 & 7.54 & 10.02 & 0.14 & 3 & 2 & 2 & 2 & 0 \\
91 & 0.0238 & 4 & 2.48 & 0.030 & 1.72 & 0.24 & -1.66 & 12.2 & 10.42 & 8.54 & 10.41 & 0.14 & 3 & 3 & 3 & 1 & 0 \\
92 & 0.0215 & 4 & 2.83 & 0.008 & 1.45 & 0.21 & -2.27 & 12.63 & 10.29 & 8.21 & 10.50 & 0.14 & 1 & 2 & 1 & 3 & 0 \\
93 & 0.0168 & 4 & 2.55 & 0.022 & 1.85 & 0.14 & -1.59 & 12.63 & 9.62 & 8.07 & 10.17 & 0.14 & 1 & 2 & 1 & 2 & 0 \\
94 & 0.0417 & 7 & 2.92 & 0.016 & 1.76 & 0.23 & -2.05 & 13.13 & 10.28 & 9.24 & 10.58 & 0.17 & 6 & 1 & 3 & 3 & 0 \\
95 & 0.0396 & 4 & 2.72 & 0.033 & 1.48 & 0.15 & -2.13 & 12.57 & 9.88 & 8.38 & 10.48 & 0.13 & 1 & 3 & 3 & 1 & 0 \\
96 & 0.0292 & 4 & 2.34 & 0.049 & 1.48 & 0.19 & -1.76 & 11.82 & 9.97 & 8.37 & 10.09 & 0.16 & 1 & 3 & 3 & 1 & 0 \\
97 & 0.0218 & 5 & 2.8 & 0.014 & 1.8 & 0.17 & -1.9 & 12.99 & 9.69 & 8.31 & 10.37 & 0.13 & 3 & 2 & 0 & 2 & 2 \\
99 & 0.029 & 5 & 2.65 & 0.025 & 1.63 & 0.14 & -1.91 & 12.51 & 9.37 & 8.52 & 10.25 & 0.14 & 2 & 2 & 1 & 3 & 1 \\ \hline
\end{longtable}
\end{landscape}

\label{lastpage}
\newpage

\end{document}